# Phenotype discovery of traumatic brain injury segmentations from heterogeneous multi-site data

Adam M. Saunders[*a], Michael E. Kim[b], Gaurav Rudravaram[a], Lucas W. Remedios[b], Chloe Cho[c], Elyssa M. McMaster[a], Daniel R. Gillis[d], Yihao Liu[a], Lianrui Zuo[a], Bennett A. Landman[†a-c,e], Tonia S. Rex[†e,f]

[a]Department of Electrical and Computer Engineering, Vanderbilt University, Nashville, TN, USA; [b]Department of Computer Science, Vanderbilt University, Nashville, TN, USA; [c]Department of Biomedical Engineering, Vanderbilt University, Nashville, TN, USA; [d]Publicis Sapient, Arlington, VA, USA; [e]Department of Radiology and Radiological Sciences, Vanderbilt University Medical Center, Nashville, TN, USA; [f]Department of Ophthalmology and Visual Sciences, Vanderbilt University Medical Center, Nashville, TN, USA

## ABSTRACT

Traumatic brain injury (TBI) is intrinsically heterogeneous, and typical clinical outcome measures like the Glasgow Coma Scale complicate this diversity. The large variability in severity and patient outcomes render it difficult to link structural damage to functional deficits. The Federal Interagency Traumatic Brain Injury Research (FITBIR) repository contains large-scale multi-site magnetic resonance imaging data of varying resolutions and acquisition parameters (25 shared studies with 7,693 sessions that have age, sex and TBI status defined – 5,811 TBI and 1,882 controls). To reveal shared pathways of injury of TBI through imaging, we analyzed T1-weighted images from these sessions by first harmonizing to a local dataset and segmenting 132 regions of interest (ROIs) in the brain. After running quality assurance, calculating the volumes of the ROIs, and removing outliers, we calculated the z-scores of volumes for all participants relative to the mean and standard deviation of the controls. We regressed out sex, age, and total brain volume with a multivariate linear regression, and we found significant differences in 37 ROIs between subjects with TBI and controls ($p < 0.05$ with independent $t$-tests with false discovery rate correction). We found that differences originated in 1) the brainstem, occipital pole and structures posterior to the orbit, 2) subcortical gray matter and insular cortex, and 3) cerebral and cerebellar white matter using independent component analysis and clustering the component loadings of those with TBI.

## 1. INTRODUCTION

Traumatic brain injury (TBI) refers to a spectrum of injuries with large variation in terms of reporting, terminology, and outcomes.[1] Traditionally, clinicians grade the severity of TBI using the Glasgow Coma Scale (GCS), which categorizes TBI into mild, moderate, and severe based on motor, verbal, and eye opening responses.[2] The GCS is useful for triaging acute TBI, but it can lead to an overly simplistic representation of TBI that does not meaningfully represent patient outcomes. For example, patients in a chronic state of mild TBI have changes in the visual system independent of reported visual symptoms.[3] In 2025, the National Institute of Neurological Disorders proposed a framework for reporting TBI based on clinical, blood biomarker, imaging, and modifiers (CBI-M) to move beyond reports of TBI solely with GCS.[4] This framework establishes TBI as a complex, heterogeneous condition compounded by diagnostics, reporting, and imaging techniques.

Phenotyping has the potential to drive understanding and disentanglement of the complex heterogeneity of TBI. Phenotyping is the process of attempting to describe subtypes or characteristics of a group, such as those with TBI. Pugh et al. (2021) compiled a review of attempts to create phenotypes for TBI and described phenotypes that cover a large variety of domains such as reported symptoms, psychological, behavioral and cognitive traits, or clinical biomarkers.[1] They identified five common phenotypes of shared outcomes from these studies: mostly healthy, mostly healthy with persistent conditions, functional issues, behavioral and mental health issues, and those with mixed traits. Nielson et al. (2017) used topological data analysis to subtype TBI based on clinical information, genetics, neuroimaging, and blood biomarkers.[5] They found a subtype of subjects with mild TBI who had normal imaging but poor outcomes with post-traumatic stress disorder at the three- to six-month follow-up. This group also had a common enriched blood biomarker.

Rosario et al. (2018) analyzed pediatric data for outcomes after severe TBI and found several clinical symptoms associated with mortality like midline shift, subarachnoid hemorrhage, high spine abbreviated injury scale, and two non-reactive pupils.[6] Changes in brain volume detectable through magnetic resonance imaging (MRI) can occur in many regions of the brain. The effect sizes of these differences are subtle due to the large variability in brain volumes and heterogeneity of injuries, especially in mild TBI.[7] Some of these changes are only detectable at follow-up visits, and Zhou et al. (2013) demonstrated small volumetric changes (less than 1 cm$^3$) in mild TBI that are associated with anxiety and depression.[8]

Due to the heterogeneity of TBI, previous works have highlighted the need for analyzing large datasets to build TBI phenotypes.[5,6] The Federal Interagency Traumatic Brain Injury Research (FITBIR) repository stores large-scale multi-site data from TBI studies across many domains like clinical symptoms and assessments, biomarkers, neuroimaging, and demographics. Studies share common data elements and global unique identifiers that can link subjects across studies.[9] As of July 2025, FITBIR contains 137,536 subjects across 157 studies, with 195,555 radiological image volumes.[10]

Large-scale MRI from multiple sites is prone to site effects and differences in acquisition parameters that are a confounding factor in analysis.[11] One solution is image harmonization, where a generative model synthesizes MRI with the desired contrast of a target image while keeping the anatomy of the original image.[12] Harmonization with attention-based contrast, anatomy, and artifact awareness (HACA3) is an image harmonization algorithm that disentangles anatomy, artifact, and contrast information from a set of images and fuses this information to harmonize to a target contrast.[13,14] HACA3 is a multi-contrast model that uses an attention mechanism to fuse information from multiple contrasts such as T1-weighted, T2-weighted and fluid-attenuated inversion recovery (FLAIR) images, and it allows for imputation of missing or imperfectly acquired contrasts. HACA3 requires no prior knowledge of site or contrast. Another way to account for differences in acquisition is performing super-resolution on images with low through-plane resolution compared to the in-plane resolution. Synthetic multi-orientation resolution enhancement (SMORE) is a self-supervised super-resolution algorithm that trains on each image individually, learning the correspondence between the low-resolution through-plane slices and the high-resolution in-plane slices to produce images with isotropic resolution.[15,16]

Past efforts of data harmonization from FITBIR have focused on non-imaging data or fields derived from radiological reports. In a series of proof-of-concept studies, O'Neil et al. harmonized FITBIR data on cognitive function,[17] post-traumatic stress disorder,[18] and depression and suicidality[19] to study relationships with TBI. Due to the enormous effort in harmonizing across different data fields and limited demographic information in common, these works focused on a few studies at a time. Yaseen et al. (2023) harmonized and imputed physiological data across time from five FITBIR studies.[20] The harmonization methods in these studies included identifying common data fields, remapping similar measures across studies, or regressing out site as a covariate. The harmonization occurred on data fields, not directly on images.

While these works highlight the usefulness of harmonizing across heterogeneous data sources from FITBIR, previous works have not addressed the need to analyze large-scale multi-site imaging data from FITBIR. Image harmonization would allow us to learn directly from images across diverse data sources. We could develop phenotypes by investigating brain volumes using segmentation algorithms rather than relying on pre-derived information such as radiology reports. In this work, we propose a pipeline for analysis of FITBIR imaging data. We harmonized and processed structural MRI from 25 FITBIR studies (out of 25 total studies with shared structural MRI available at the time of download), enabling the development of phenotypes of differences in brain volume in those with TBI compared to controls.

## 2. METHODS

**Standardized organization of FITBIR imaging data**

Here, we used MRI from FITBIR to investigate common signatures of TBI in neuroimaging. We included data with shared structural MRI (i.e., the data were not private at the time of download) with a T1-weighted, T2-weighted, or T2-weighted FLAIR contrast ($n = 26$). We excluded studies with only a single MRI image shared ($n = 1$). In total, 25 studies met the criteria. Available MRI were stored mostly in DICOM format with some in NIfTI format and heterogeneous data organization between studies.

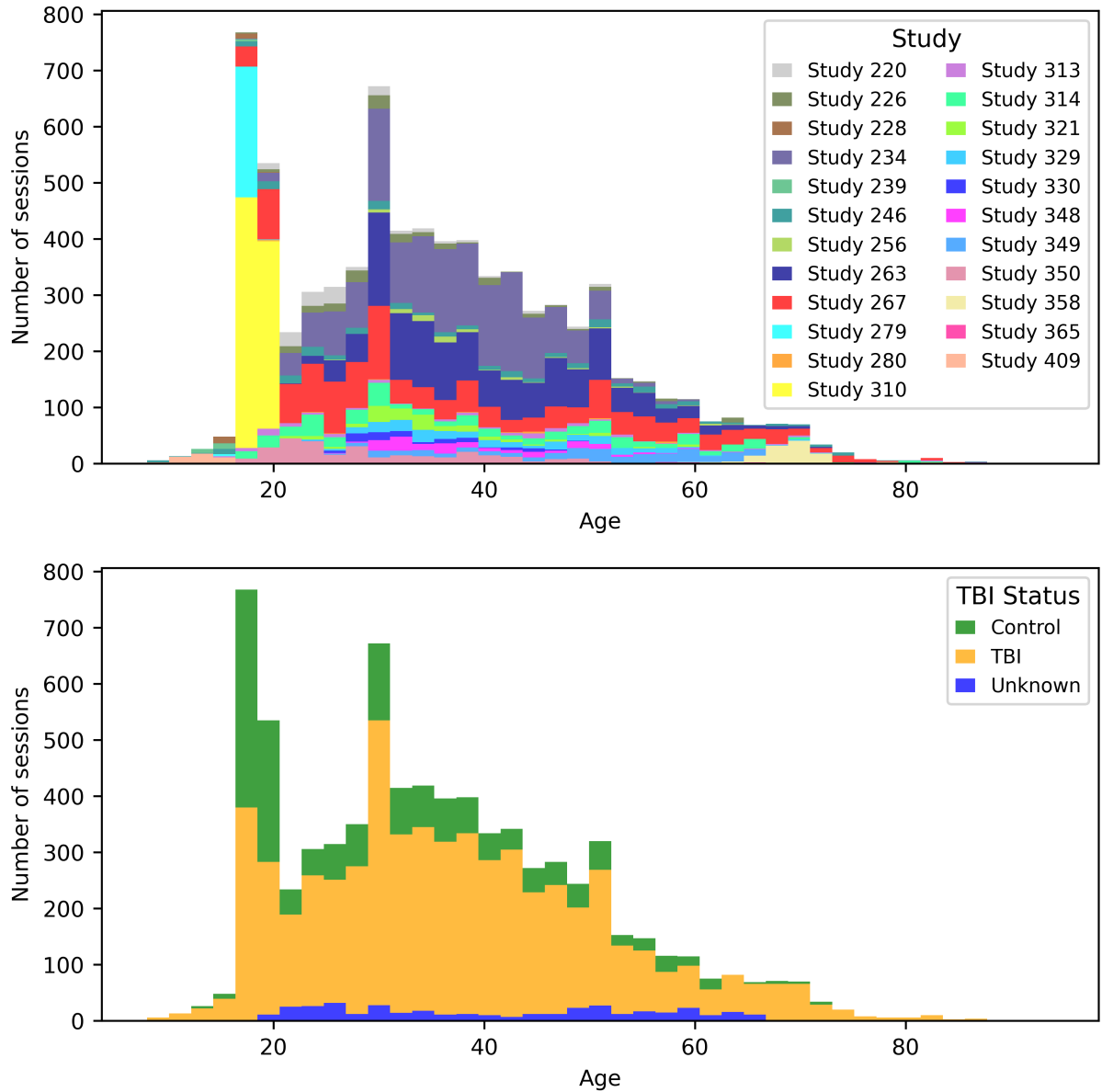

Figure 1. The FITBIR repository contains large-scale multi-site data from subjects with TBI and controls from a large range of ages. Across 25 studies, there are 8,212 MRI sessions. 1,882 sessions are from controls, 5,811 are from those with TBI, and 519 have an unknown TBI status. Only sessions where age is available are shown.

We converted all DICOM data to NIfTI format using dcm2niix[21] and reorganized the studies into the brain imaging data structure (BIDS) format for consistency.[22] We aggregated subject demographics such as age, sex, race, ethnicity, and TBI status as available. We treated TBI status as a binary variable, counting subjects in the case group of the study or those with a recorded GCS as subjects with TBI. Some of the studies included only a case or control group, and some studies only included those with mild TBI. We then ran quality assurance (QA) on the MRI by visually inspecting a montage of slices using AutoQA.[23] We flagged any images with extremely limited fields of view, severe motion or imaging artifacts, sessions acquired with a contrast agent, and severe trauma or missing brain tissue that would render automated segmentation methods impractical. After initial QA, we had 5,600 subjects across 8,212 imaging sessions with a total of 29,184 images across a large variety of ages (Fig. 1). Some of the subjects had multiple sessions.

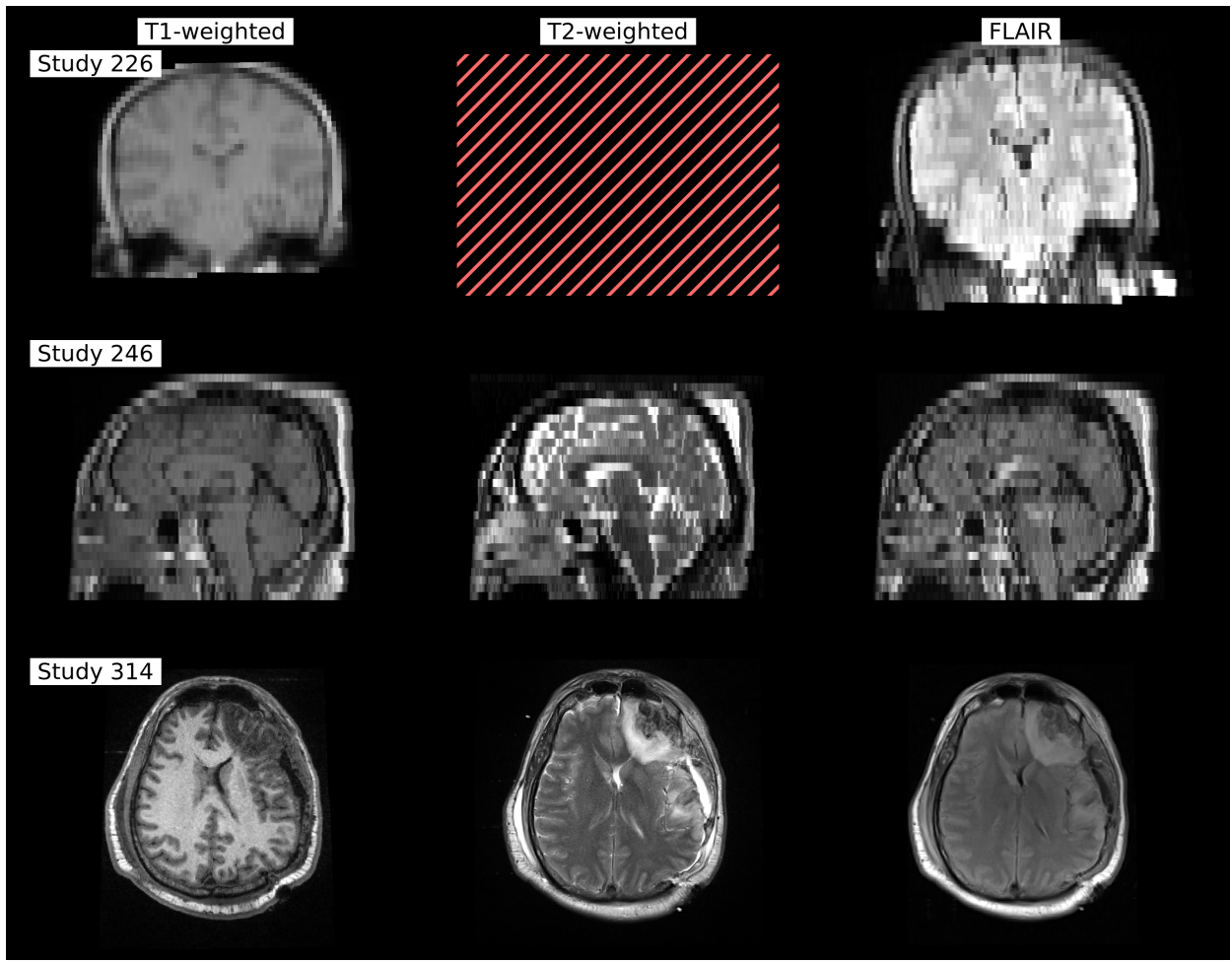

Figure 2. Images from three example studies in the FITBIR repository demonstrate the heterogeneous presentation in structural MRI data. The FITBIR repository contains imaging data from studies with highly variable acquisition parameters and contrast. Some contrasts are missing, such as the T2-weighted image from Study 226. The resolutions vary across the through-plane and in-plane of the images. We show sagittal and coronal views from Studies 226 and 246 to demonstrate through-plane slices. Despite the low resolutions and highly variable anatomy, these three images passed quality assurance.

**Image preprocessing and harmonization**

The images varied widely in terms of resolution, contrast and anatomy based on the study, which could be a confounding factor in our analysis (Fig. 2). We used image harmonization to standardize contrast and resolution across the heterogeneous multi-site data. We chose to harmonize the images to a private dataset acquired locally at Vanderbilt University Medical Center with 63 subjects using HACA3. For this dataset, we acquired T1-weighted and T2-weighted images after obtaining informed consent under a protocol approved by the institutional review board.[3] An automated tool converted the data to BIDS format from an institutional medical imaging archive.[24] For all of the datasets, we chose a single T1-weighted, T2-weighted and FLAIR image (if available) to represent each session. For sessions where multiple scans are available for a given contrast, we select the image with the higher resolution.

We preprocessed the images as described in the original HACA3 paper by Zuo et al.[13] by first applying SMORE for super-resolution on images that have a voxel size along one dimension more than three times the size of the smallest voxel size. After applying super-resolution, we applied bias field correction with N4.[25] Finally, we registered all the images to a common space using the International Consortium for Brain Mapping 2009c nonlinear symmetric Montreal Neurological

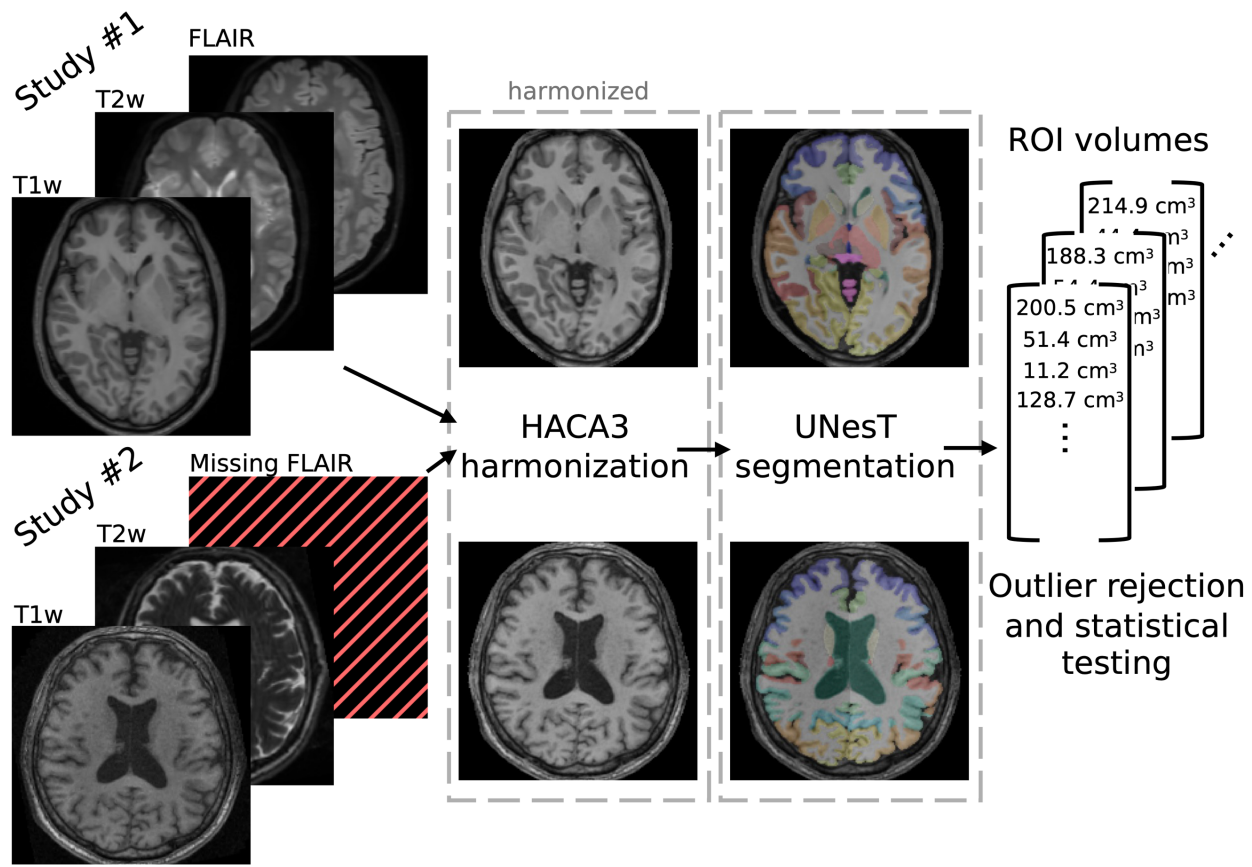

Figure 3. Using HACA3, we harmonized data from each session to a T1-weighted image from a locally acquired dataset. We applied the UNesT segmentation algorithm to calculate the volumes of regions of interest (ROI) in the brain. We standardized the volumes relative to the controls and rejected outliers based on Mahalanobis distance greater than 500, indicating a very extreme outlier.

Institute (MNI) template.[26] Intra-session registration is a key component for HACA3. Thus, we first registered all the contrasts to a T1-weighted image from the session (if available), then registered the T1-weighted image to the MNI template. Finally, we applied this transformation to the other contrasts. If a T1-weighted image was not available for a session, we directly registered all the available contrasts to the MNI template. We also ran QA as above after preprocessing to ensure the data were registered correctly.

For training the MRI harmonization model HACA3, we randomly sampled 12 sessions from each study (10 for training, 2 for validation), prioritizing sessions that had the most available contrasts. No subjects or sessions were shared between the training and validation set. We fine-tuned the publicly available weights trained by Zuo et al. and trained with the same hyperparameters and anatomy, contrast, and artifact encoders as in Zuo et al.[13] We stopped training when the validation loss had not decreased in 50 passes of the validation set. After training, we harmonized using all the available contrasts from each session to a randomly selected T1-weighted image from the locally acquired dataset.

**Brain segmentation and phenotype discovery**

Using the preprocessed images, we aimed to create phenotypes or signatures of TBI from brain volumes. To segment the brain tissue into 132 regions of interest (ROIs), we used the UNesT algorithm to segment the T1-weighted images.[27] After running QA on the resulting segmentations, we calculated the volumes of each ROI for each image. Treating each session as independent, we z-scored the volumes of each ROI relative to the mean and standard deviation of the controls and removed any outlier images with a Mahalanobis distance larger than 500, which is an extreme outlier. Malanobis distance measures the distance from the subject's z-score from the distribution of all z-scores. We checked the distributions of the brain volumes to confirm they were all approximately Gaussian. We regressed out age, sex, and total brain volume. Demographic information for age, sex, and TBI status were only available in subjects from 7,693 sessions (5,811 with TBI and 1,882 controls). To characterize group-level changes in brain volume, we ran an independent $t$-test comparing the volumes of subjects with TBI and control subjects across the 132 ROIs (excluding background). To correct for multiple comparisons, we used Benjamini-Hochberg false discovery rate correction ($\alpha = 0.05$).[28] Fig. 3 shows an overview of the group-level analysis.

The group-level differences between controls and those with TBI could originate from several sources or signatures of TBI relative to controls. To further investigate the sources of difference in brain volumes, we ran an independent component analysis (ICA) on the $z$-scores of the controls to generate 23 independent components (95% explained variance). Then, we calculated the subject loadings of the $z$-scores of those with TBI onto these components. These loadings represented how these 23 independent sources of variation in brain volume in normal brains differ compared to those with TBI. We ran $k$-means clustering to generate three clusters of TBI loadings that represent three phenotypes of differences in TBI. We chose three clusters because we found that this number resulted in more consistent clusters across random initializations than higher numbers of clusters.

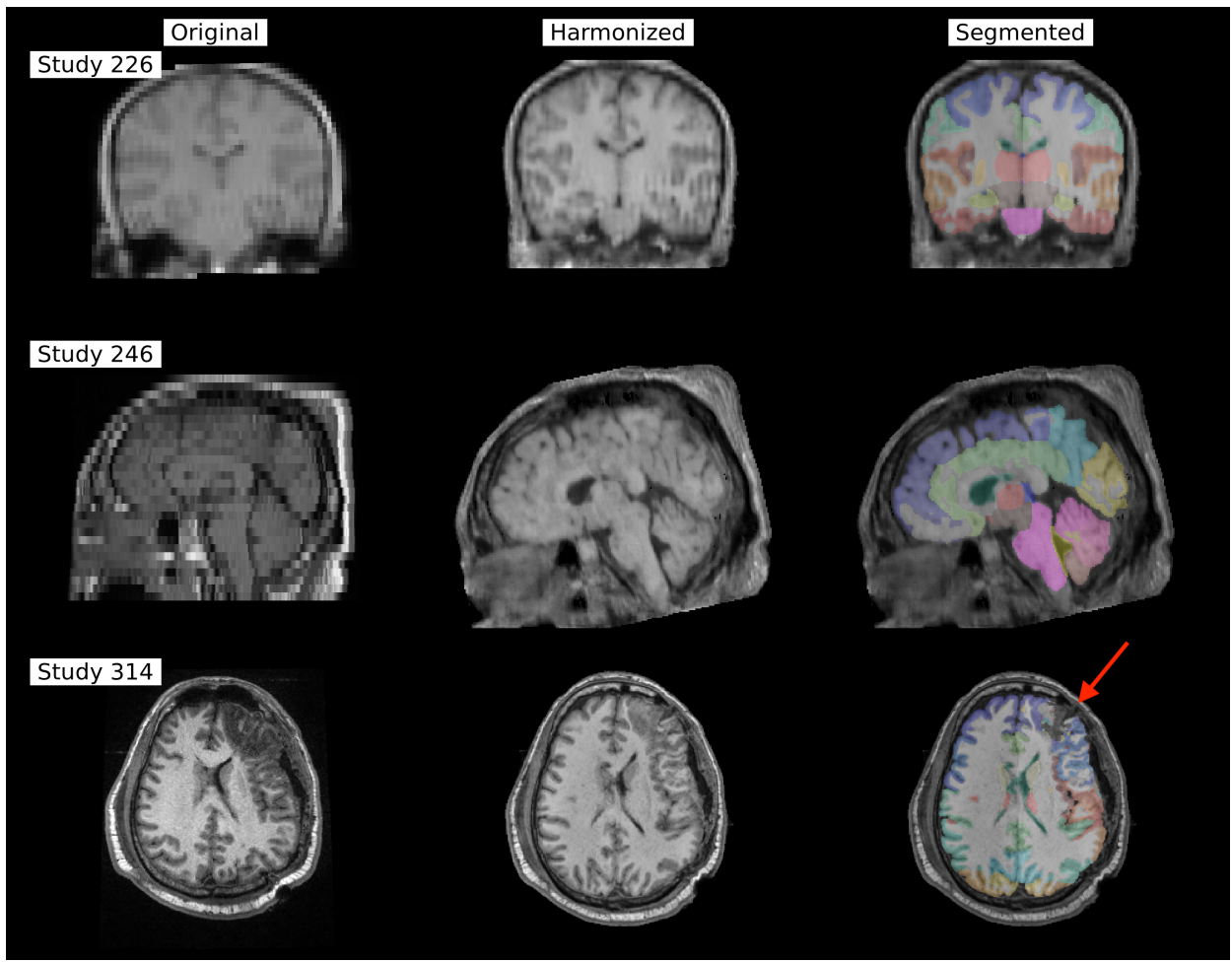

Figure 4. After applying our preprocessing pipeline using super-resolution and image harmonization, the same three T1-weighted images shown in Fig. 2 appear much more standardized in resolution and contrast. The low through-plane images from Studies 226 and 246 have recovered some details through super-resolution. Despite the variable anatomy and poor gray matter-white matter contrast in the original images, the UNesT segmentation algorithm is fairly robust on the ROIs, failing in regions of high anatomical deformation in Study 314 (see red arrow).

# 3. RESULTS

After applying our preprocessing involving super-resolution and image harmonization, the resulting set of synthesized T1-weighted images was more consistent in anatomical contrast and resolution (Fig. 4). The gray matter-white matter contrast was enhanced, despite the poor contrast available in the T1-weighted images for Studies 226 and 246. The UNesT brain segmentation algorithm segmented the ROIs well, though failing in regions of moderate anatomical deformations present in the original images. After calculating the volumes of each ROI and *z*-scoring relative to the controls, we identified and removed 47 outlier sessions whose *z*-scores had a Mahalanobis distance larger than 500. All the outliers came from sessions from subjects with TBI. While the outlier images all passed QA, most had large anatomical deformations resulting in large volumes for certain ROIs like the ventricles.

We identified 37 ROIs with a significant difference in *z*-scored brain volumes after false discovery rate correction (Fig. 5). The effect sizes were all mild; the largest magnitude of effect size was 0.26. Most of the effect sizes were negative, indicating a decrease in brain volume in subjects with TBI. However, several of the ROIs had a positive effect size corresponding to an increase in brain volume in subjects with TBI, namely the areas posterior to the orbit of the eye like the subcallosal area, gyrus rectus, orbital gyrus, and occipital pole.

Significant ROIs are concentrated in specific regions of the brain (Fig. 6). The temporal and occipital poles show significant changes, as well as many structures around the orbit of the eye. The white matter and brainstem also showed significant differences, as well as areas in the cerebellum. The significant regions were, for the most part, symmetric across the brain.

Next, after running our phenotype generation by generating independent components and clustering the loadings of subjects with TBI, we generated three regional brain volume signatures of differences in those with TBI (Fig. 7). Cluster 1 showed changes in the occipital pole, temporal pole, structures posterior to the orbit of the eye, and brainstem compared to controls. Cluster 2 mainly demonstrated changes in subcortical gray matter like the pallidum and putamen and areas in the insular cortex. Finally, Cluster 3 had changes mainly in the cerebral and cerebellar white matter. While the signs of the ICA loadings are arbitrary and do not represent increases or decreases in brain volume, we can interpret the relative signs within each cluster. Clusters with inverse relationships indicate one area increased in volume and another area decreased. Cluster 1 showed inverse changes in the structures around the orbit and temporal pole versus the brainstem, whereas Cluster 2 showed an inverse relationship between changes in the subcortical gray matter and insular cortex versus the brainstem.

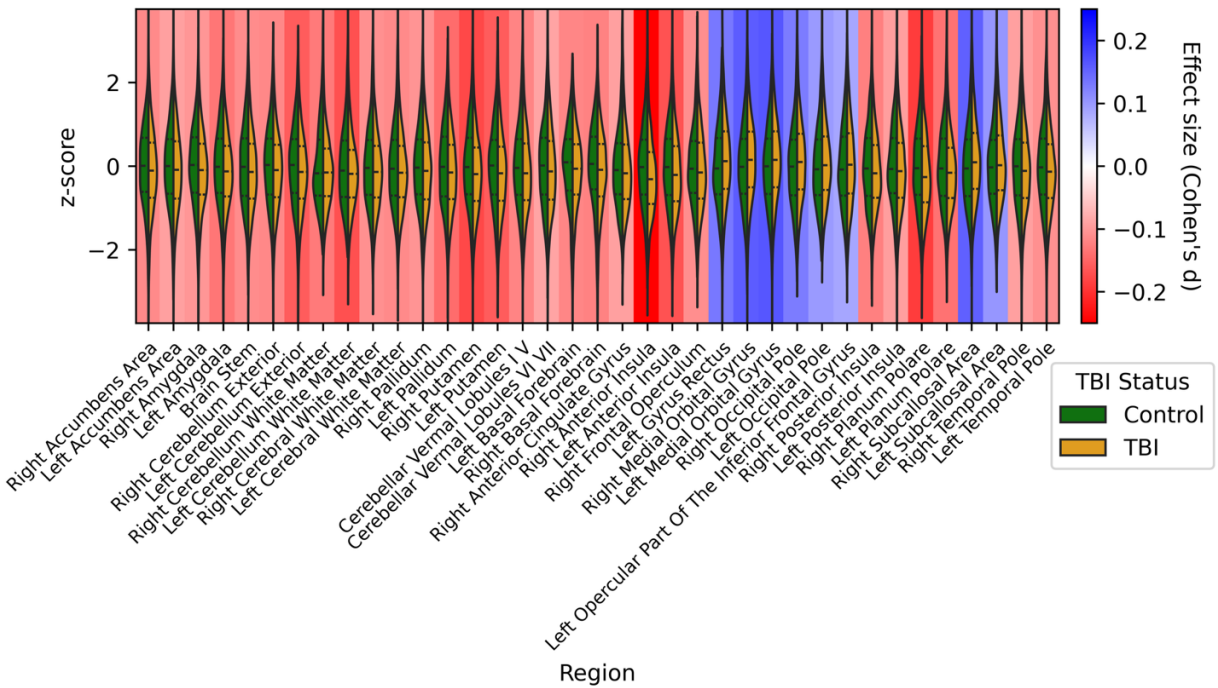

Figure 5. We identified 37 ROIs with significant differences in volume in TBI compared to controls (independent t-test, $\alpha = 0.05$ with false discovery rate correction). Each difference had a mild effect size, measured with Cohen's $d$, and most ROIs had a decrease in brain volume in TBI compared to controls.

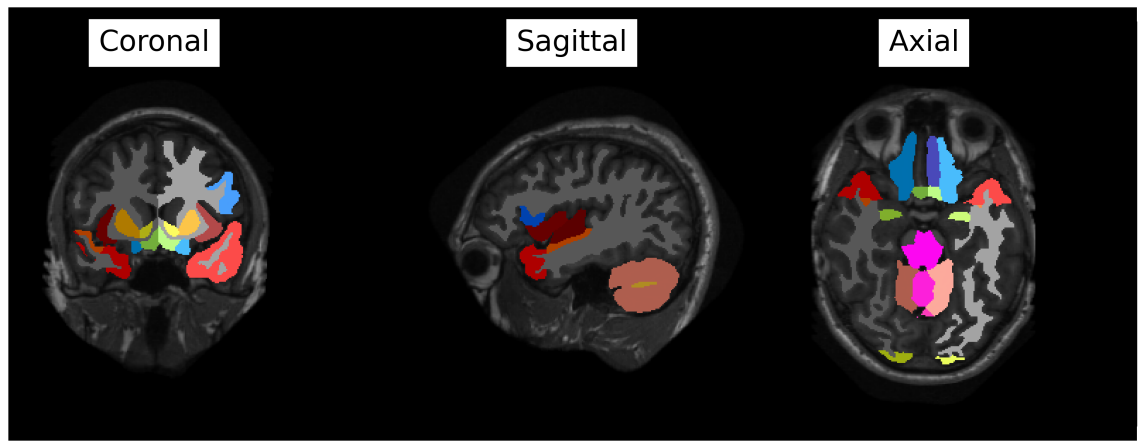

Figure 6. The 37 ROIs with significant differences involve the white matter, cerebellum, brainstem, and structures around the orbit of the eye, as well as the occipital poles. The significant differences appear to be largely symmetric across the two halves of the brain, and areas of change concentrate in the areas near the orbit of the eye. Additionally, the temporal pole and occipital pole are significantly different.

## 4. DISCUSSION

The changes in brain volume in TBI can be very subtle, especially with mild TBI.[7] To detect nuanced changes with small effect sizes, we need high statistical power. Therefore, it is necessary to perform these studies with as many subjects as possible. Using a large-scale, multi-site dataset allowed us to identify small but significant differences in brain volume between controls and subjects with TBI that would be difficult to detect in smaller datasets.

Several of the identified ROIs overlap with previous research. For example, Bigler mentions decreases in the basal ganglia, cerebellum and increases in the ventricular volumes,[7] changes also detected here. However, many of the ROIs we identified as significant lie outside of the "cone of vulnerability" in the midbrain.[7] The ROIs appear clustered by physical location in the brain. Many of the ROIs we identified are adjacent to the bony structures of the orbit of the eye. The occipital pole near the posterior extreme of the brain was also affected. Further study is warranted to connect these ROIs and identified phenotypes with underlying mechanisms of injury. Additionally, we generated independent components from the controls and then projected those with TBI onto these components, meaning our phenotypes are relative to the independent sources of variation in controls. We could have generated the phenotypes on independent components derived only from those with TBI if we instead wanted to analyze sources of variation in those with TBI.

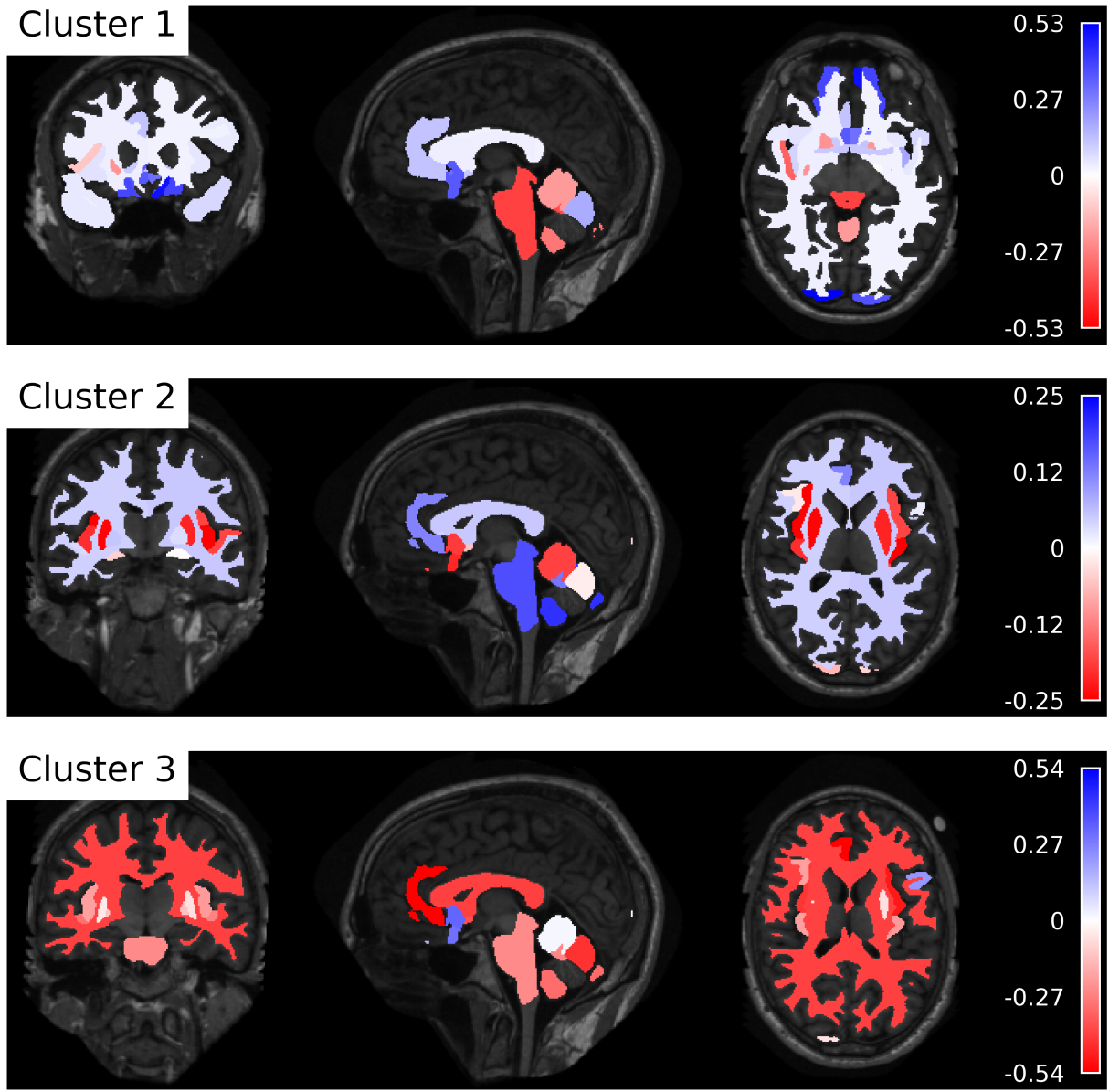

Figure 7. We ran independent component analysis on the z-scores of the controls to generate 23 components (95% explained variance). After projecting those with TBI onto the independent components, we clustered the loadings to generate three signatures of changes in volume in those with TBI. These clusters showed changes in 1) the brainstem, occipital pole and structures posterior to the orbit, 2) subcortical gray matter and insular cortex, and 3) cerebral and cerebellar white matter. The color bars indicate the magnitude and sign of each ROIs contribution to the cluster. Note the signs of the loadings are arbitrary and do not necessarily correspond to the signs of changes in brain volume, but Clusters 1 and 2 show inverse relationships where one area increased in volume and another decreased in volume.

Analysis of multi-site data from the FITBIR requires careful preprocessing and consideration of the differences between studies. For imaging data, super-resolution and harmonization allows for accounting for acquisition differences among sites. We chose to synthesize and analyze information from T1-weighted images to use well-validated automated segmentation tools. Future work could synthesize any desired contrast with HACA3. SMORE helped super-resolve some details in images with a poor through-plane resolution. In this analysis, we treated sessions as independent, but several of the studies contained longitudinal data from the same subjects. We analyzed only data that had age, sex, and TBI status defined to allow for regressing out these factors. These demographics are not available for all data, and previous studies have also limited their analysis to data in FITBIR with relevant demographics available.[17,18,20] Also, the demographics of the studies may have biased the results without matching samples from the controls and those with TBI. For example, many of the controls were from younger subjects.

## 5. CONCLUSION

We presented a method for discovering phenotypes from TBI imaging data acquired under different acquisition parameters and resolutions. Using a consistent imaging data structure and preprocessing including super-resolution and image harmonization allowed us to compare brain volumes across 25 heterogeneous FITBIR studies. We identified 37 ROIs with significant differences between controls and those with TBI, and we identified three phenotypes from these differences. Additionally, many of the identified regions were physically located near the orbit of the eye. This study demonstrates the feasibility of analyzing TBI imaging from large-scale multi-site data, developing phenotypes from 25 studies with 7,693 imaging sessions.

## ACKNOWLEDGEMENTS

This work was supported by DoD grant HT94252410563, National Cancer Institute grants R01CA253923 and R01CA275015, NIH 5U01DA055347-03, R01HL169944, U24AG074855, R01MH121620, NIH NIGMS T32GM007347, and Integrated Training in Engineering and Diabetes grant number T32 DK101003. This work was supported by the Alzheimer's Disease Sequencing Project Phenotype Harmonization Consortium (ADSP-PHC) that is funded by NIA (U24 AG074855, U01 AG068057 and R01 AG059716). This work was conducted in part using the resources of the Advanced Computing Center for Research and Education at Vanderbilt University, Nashville, TN. The Vanderbilt Institute for Clinical and Translational Research (VICTR) is funded by the National Center for Advancing Translational Sciences (NCATS) Clinical Translational Science Award (CTSA) Program, Award Number 5UL1TR002243-03. Data and/or research tools used in the preparation of this manuscript were obtained and analyzed from the controlled access datasets distributed from the DOD- and NIH-supported Federal Interagency Traumatic Brain Injury Research (FITBIR) Informatics Systems. FITBIR is a collaborative biomedical informatics system created by the Department of Defense and the National Institutes of Health to provide a national resource to support and accelerate research in TBI. Dataset identifiers: FITBIR-STUDY0000220, FITBIR-STUDY0000226, FITBIR-STUDY0000228, FITBIR-STUDY0000234, FITBIR-STUDY0000239, FITBIR-STUDY0000246, FITBIR-STUDY0000256, FITBIR-STUDY0000263, FITBIR-STUDY0000267, FITBIR-STUDY0000279, FITBIR-STUDY0000280, FITBIR-STUDY0000310, FITBIR-STUDY0000313, FITBIR-STUDY0000314, FITBIR-STUDY0000321, FITBIR-STUDY0000329, FITBIR-STUDY0000330, FITBIR-STUDY0000348, FITBIR-STUDY0000349, FITBIR-STUDY0000350, FITBIR-STUDY0000358, FITBIR-STUDY0000363, FITBIR-STUDY0000364, and FITBIR-STUDY0000409. This manuscript reflects the views of the authors and may not reflect the opinions or views of the DOD, NIH, or of the Submitters submitting original data to FITBIR Informatics System.

We used generative AI to create code segments based on task descriptions, as well as to debug, edit, and autocomplete code. Additionally, generative AI technologies have been employed to assist in structuring sentences and performing grammatical checks. The conceptualization, ideation, and all prompts provided to the AI originated entirely from the authors' creative and intellectual efforts. We take accountability for the review of all content generated by AI in this work.